\documentstyle[12pt,amssymb]{article}
\begin{document}
\thispagestyle{empty}
\vglue 1cm
\begin{flushright}
CERN--TH/2000--146
\end{flushright}
\vglue.5cm
\begin{center}
{\bf THE SO-CALLED RENORMALIZATION GROUP METHOD APPLIED} \\
{\bf TO THE SPECIFIC PRIME
 NUMBERS LOGARITHMIC DECREASE}\\
\vglue1cm
A. Petermann\\ 
TH Division, CERN\\
CH - 1211 Geneva 23\\

\vglue2cm
ABSTRACT 
\end{center}
 A so-called Renormalization Group (RG) analysis is performed in
order to shed some light
 on why the density of prime numbers in $\Bbb N^*$ decreases like the
single power of the inverse neperian logarithm.

\vglue3cm
\begin{flushleft}
CERN--TH/2000--146\\
May 2000
\end{flushleft}

\newpage
\setcounter{page}{1} 

\part{ }
\noindent
{\Large \bf The most elementary proof of the Prime Numbers Theorem}
\vglue1cm

These few lines are not part of the proof. They simply show the history which has
led to the starting point of our proof. The two main steps of it involve  only
formal elementary algebra, with no recourse naturally to Functions' theory nor
complex variables. Euler proved in 1747 \cite{peter1}, quite formally, that the
prime numbers are linked with natural integers, by establishing what is universally
known as ``Euler identity'' and which is so famous that we do not recall it here.

From this identity, one can deduce straight forwardly an approximate
formula\footnote{Which, in fact, is the strict equivalent of Euler identities.}
giving 
$$
{\rm (I)}\sum_{p < \Lambda} 1/p = 1 \cdot \log (\log \Lambda ) + \cdots\;. \label{peter:eq1}
$$
This result has been refined, mainly by Mertens \cite{refpeter5} with the aim to
establish the value of constants possibly entering a much more exact expression of
the sum in (I). This expression, we call it ``Euler-Mertens identity''
and is the starting point of our proof, the formula \ref{peter:eq1} of our theorem.
\vglue.5cm

\underline{\bf Proof}
\vglue.5cm
Consider the Euler-Mertens identity (I), exact when $\Lambda \rightarrow
\infty$. We introduce it under the following form (\ref{peter:eq1}) and apply the RG
analysis \cite{refpeter1}
\begin{equation}
\bar F = (\log \log \Lambda)^{_1} \int^\Lambda \frac{dn \, \bar n (n)}{n} \cong
C^{te} \cong 1 \, , \;\; \Lambda \rightarrow \infty\label{peter:eq1}
\end{equation}
$\bar n(n) dn$ is a measure $df (n)$ of the Stieltjes type and $\bar n (n)$ can be
considered as a density in the physical sense\footnote{$df (n) = \bar n (n) dn$ and,
according to the result (\ref{peter:eq3}) $f (\Lambda ) =\int^\Lambda \frac{dn}{\log
n} \cong Li (\Lambda)$.}.

Then , since
\begin{eqnarray*}
\Lambda \, \partial/\partial\Lambda \bar F &=& \frac{\partial } {\partial \log
\Lambda} (\log \log\Lambda)^{-1} \cdot \int^\Lambda \frac{dn \bar n(n)}{n}
+\frac{\bar n (\Lambda)}{\log\log \Lambda}\\
&&\cr
&=&-(\log \Lambda)^{-1} (\log\log \Lambda)^{-2} \cdot \int^\Lambda \frac{dn \bar
n(n)}{n} + \frac{\bar n (\Lambda)}{\log\log\Lambda}
\end{eqnarray*}
one deduces, by the RG method, (since $d\bar F/d\log \Lambda = 0)$
\begin{eqnarray}
0&=&-\frac{(\log \Lambda)^{-1}}{\log\log\Lambda} \cdot F + \frac{\bar n
(\Lambda)}{\log \log \Lambda} + \left[{\frac{\partial \bar n (\Lambda)}{\partial
\log \Lambda}}\right] \frac{\delta } {\delta \bar n (\Lambda)} \bar F\cr
&&\cr
&=& \left[{ - (\log \Lambda)^{-1} + \bar n (\Lambda)}\right] (\log \log \Lambda)^{-1}
+ 0 \left({\frac{1}{\Lambda \cdot \log^2\Lambda \cdot \log \log \Lambda}}\right)\;.\label{peter:eq2}
\end{eqnarray}
So from \ref{peter:eq2}, at this approximation one gets 
\begin{equation}
\bar n (\Lambda) \cong (\log \Lambda )^{-1} \;, \;\;\; \Lambda \gg 1\label{peter:eq3}
\end{equation}
which is the prime Numbers theorem, since the density 
$$ \bar n (\Lambda) = \Lambda^{-1} \cdot \pi (\Lambda)$$

\part{ }
\noindent
{\Large \bf The RG Equation for the Density of Prime Numbers}\footnote{The RG-method
has been designed in order to know how the structure of a theory gets modified when
the scale is changed.}
\vglue1cm

The density of natural integers is scale invariant:
\begin{equation}
\lambda \frac{\partial}{\partial \lambda} \bar{d} (n_i \lambda , \bar{d} (1))
=0~,
\label{eq:1}
\end{equation}
$\bar {d}$ being the density around $n_i\lambda$, and $\lambda 
\frac{\partial}{\partial \lambda}$ the generator of scale transformations of the
natural integers.

If, on the other hand, $n_i$ is a prime and $\lambda n_i$ is around another prime,
say $n^\prime_i$, then

$$\bar d (n_i^\prime \cong \lambda n_i , \bar d (1)) \neq \bar d (n_i, \bar d(1))\;
.$$

In such a case, instead of having an equation like (\ref{eq:1}), expressing the
invariance for scale changes,  one uses, as is customary, the so-called
Renormalization Group (RG) equation (or better: renormalization transformation
equation) which generally substitutes Eq.~(\ref{eq:1}) when scale invariance is
broken. The strategy is to compensate the broken invariance, for example in the
present case, by a density $\bar d$, which this time depends upon $\lambda$ and is
different from that of (\ref{eq:1}), namely $\bar d (1)$. The RG equation, as is
well known
\cite{refpeter1}, reads

\begin{equation}
\lambda \frac{\partial}{\partial \lambda} \bar d (n_i\lambda , \bar d (\lambda)) +
\left[{\frac{\lambda\partial} {\partial\lambda} \bar d (\lambda)}\right] \frac{\partial}{\partial
\bar d (\lambda)} \bar d (n_i \lambda , \bar d(\lambda)) = 0\;.
\label{eq:2}
\end{equation} 
Equation (\ref{eq:2})\footnote{\mbox{$n_i\lambda \cong$ prime, as $n_i$ is.}}
introduces the quantity
$[\lambda \frac{\partial}{\partial\lambda} \bar d (\lambda)] \frac{\partial}
{\partial
\overline{ d }(\lambda)}$ which is a one-dimensional vector field ${\underline{V}} (\bar
d(\lambda))$ on the axis of integer numbers.

Now, the problem is to solve (\ref{eq:2}) for $\bar d$.

This solution, as  will be explained in the Appendix, when taken
between  two different numbers $N_1$ and $N_2$, turns out to be
\begin{equation}
\frac{1}{\bar d (N_1)} - \frac{1}{\bar d(N_2)} = \log \frac{N_1}{N_2}\; .
\label{eq:3}
\end{equation} 
But, perhaps, more instructively for the scope of this short note,
\begin{equation}
\bar d (t , \bar d (0, \bar d_0) = \frac{\bar d (0, \bar d_0)}{1+ t 
\cdot \bar d (0, \bar d_0)}\label{eq:4}
\end{equation} 
with $t = \log N$, exhibiting the neperian logarithmic  single power
decrease as having 
 its  origin in the violation of the scale invariance symmetry.

Indeed, if in the region of large primes (say between $10^{15}$ and $2\times
10^{17}$),  the numerical results obtained by the use of (\ref{eq:3}), formula
(\ref{eq:3}) does not tell else that, for each number $N_1$ and $N_2$,
$$\frac{N}{\pi(N)} = \log N$$ 
which is a 100 years-old, over-demonstrated
asymptotic result
\cite{refpeter2,refpeter3}.

However, in this note, our aim is to look for the deep reason  why the 
density of primes decreases with the single power of the natural logarithm. We hope
that we have been able to shed some light on this fact: the breaking of a
symmetry, namely that of scale invariance with generator $\lambda
\frac{\partial}{\partial\lambda}$, is the very factor responsible for this specific
decrease.

The coincidence of the results obtained is striking when compared to the formulas of
the  first non-trivial approximation of Quantum ChromoDynamics (mutatis mutandis, of
course,  the concepts between two such different fields).

But a main common feature emerges: in both cases the two fields are afflicted by the
same  broken symmetry, that of scale invariance.

\newpage
\renewcommand{\theequation}{A.\arabic{equation}}
\setcounter{equation}{0}
\appendix{}
\begin{center}{\bf APPENDIX}\end{center}
\vglue1cm
\begin{itemize}
\item[1)] For natural integers, scale invariance holds for the density $\bar d (n)$, 
i.e. when $n\rightarrow \lambda n , \bar d (n) = \bar d (\lambda n)$.

\item[2)] For primes $p_i$,
$$p_i \rightarrow \lambda p_i \cong p_j , \bar d (p_j) \neq \bar d (p_i)\;,$$ so
that $\bar d$ becomes a function of $\lambda$.
\end{itemize}

One finds easily that a functional equation of the type
\begin{equation}
\bar d (\lambda n_i , \bar d (\lambda)) = \bar d (n_i , \bar d_0)\;,\label{eq:5}
\end{equation}
$(\bar d_0$ fixed, and $n_i$ representing primes as well as $\lambda n_i$ in the 
dose vicinity of $n_j$, prime itself) exists.

The RHS is $\lambda$-independent and one gets at once
\begin{equation}
\lambda \frac{\partial}{\partial\lambda} \bar d (\lambda n_i \cong n_j, 
\bar d (\lambda)) + \left[{\lambda \frac{\partial}{\partial\lambda} \bar d
(\lambda)}\right] \frac{\partial}{\partial
\bar d (\lambda)}\bar d (\lambda n_i, \bar d (\lambda ))= 0~.\label{eq:6}
\end{equation} 
Or else, calling $t = \log \lambda$ and passing to logarithmic
variables
\begin{equation}
\left[{\frac{\partial}{\partial t} + \underline{V} (\bar d (t)}\right] \bar d (t +
\log n_i , \; d (t))=0 \label{eq:7}
\end{equation} 
with 
$$\underline{V} (\bar d (t)) =\left[{ \frac{\partial}{\partial t} \bar d (t)
}\right] 
\frac{\partial}{\partial \bar d(t)}~.$$ 
To solve (A.2), one proceeds in the following way
\begin{eqnarray*}
\bar d (\delta t, \bar d (0, \bar d_0))& =& \bar d (0, \bar d(0, \bar d_0)\cr &&\\
&&+ \delta t \underline{V} (\bar d (0, \bar d_0)) \cdot \bar d (0, \bar d_0) + 0
(\delta t^2)
\end{eqnarray*} by Taylor-expanding  $\bar d$ around $\delta t =0$.

According to the properties of flows of vector fields, one has $\bar d (0, x)= x$,
that is,  for example, $\bar d (0, \bar d(0,\bar d_0))=\bar d(0, \bar d_0)= \bar
d_0, \; \bar d_0$ being a fixed arbitrary density.

One gets then
$$\bar d (\delta t , \bar d_0) - \bar d_0 = \delta t \underline{V} (\bar d_0) \cdot
\bar d_0$$ 
or\footnote{In passing we recall a well-known property of the flows $\bar
d (t, x)$;  namely they satisfy the one-parameter Abelian group: $\bar d (s+t, x) =
\bar d (t, \bar d(s,x))$, i.e. the composition law $\bar d_{s+t} = \bar d_{t} \circ
\bar d_s$. This group is trivially generated as the one-parameter group of
diffeomorphisms by the vector field $V$ on the manifold considered. For details see
Ref.~\cite{refpeter1}a.)}
\begin{equation}
\bar d (\delta t , \bar d_0) = (1+ \delta t \underline{V} (\bar d_0)) \bar d_0
\label{eq:8}
\end{equation}

By theorems by Chebyshev and Mertens \cite{refpeter4,refpeter5}, $V(x)$ can be shown
to be quadratic in its argument $x$.

It remains to exponentiate the RHS of (A.4):
\begin{eqnarray}
\bar d (t, \bar d_0) &=& \left({1 + t \underline{V} (\bar d_0) \frac{\partial}
{\partial \bar d_0} + \frac{t^2}{2!} V (\bar d_0)\frac{\partial}{\partial \bar d_0}
\cdot V (\bar d_0)
\frac{\partial}{\partial \bar d_0} +\cdots }\right)\bar d_0\cr &&\cr &=& \frac{\bar
d_0}{1+t \bar d_0}\; ;\;\;\; (V (\bar d_0) = -\bar d_0^2 \;.\;\mbox{see above})\;.
\label{eq:9}
\end{eqnarray} 
(\ref{eq:9}) is the formula (\ref{eq:4}) of the text and seems to us
 to be an explanation we were searching for,  to explain the decrease of $\bar d(t,
d_0)$ with a single power of the natural logarithm $t= \log N$ (Remember that $\bar
d(t,
\cdots) \equiv \bar d (e^t,
\cdots)$, as (\ref{eq:6}) and (\ref{eq:7}) show  without further comments.)

As a final remark, (\ref{eq:3}) follows straight forwardly from (\ref{eq:4}) by
trivial algebra.

Take (\ref{eq:4}) with two different values for $t$: $t_1 = \log N_1$, and $t_2 =
\log N_2$. It follows at once that
$$\bar d^{-1} (t_1, \bar d_0) - \bar d^{-1} (t_2, \bar d_0) = \log N_1/N_2~.$$

(Additionally it confirms the arbitrariness of $\bar d_0$ which might be chosen at
will; $\bar d_0 =1$, for example.)

\vfill\eject

\end{document}